\documentclass[5p,times]{elsarticle}
\usepackage[utf8]{inputenc}
\usepackage{graphicx}
\usepackage{listings}
\usepackage{amsmath}
\usepackage{amsfonts}
\usepackage{amssymb}
\usepackage{algorithm2e}
\usepackage{multirow}
\usepackage[hyphens]{url}
\usepackage{hyperref}
\hypersetup{breaklinks=true}
\usepackage{breqn}
\usepackage{centernot}

\journal{Future Generation Computer Systems}

\begin{document}

\begin{frontmatter}

\title{A compact QUBO encoding of computational logic formulae demonstrated on cryptography constructions}
%
%
\author{Gregory Morse$^{a,b}$} 
\ead{gmorse@groq.com}

\author{Tam\'{a}s Kozsik$^{a}$} 
\ead{kto@elte.hu}

\author{Oskar Mencer$^{b}$} 
\ead{omencer@groq.com}

\author{P\'{e}ter Rakyta$^{b,c,d}$}
\ead{prakyta@groq.com}
%

\affiliation{organization={Department of Programming Languages and Compilers, Eötvös Loránd University}, addressline={}, city={Budapest}, postcode={1117}, country={Hungary}}

\affiliation{organization={Groq Inc.}, addressline={400 Castro St Mountain View}, city={California}, postcode={94041}, country={US}}

\affiliation{organization={Department of Physics of Complex Systems, Eötvös Loránd University}, addressline={}, city={Budapest}, postcode={1117}, country={Hungary}}

\affiliation{organization={Wigner Research Center for Physics}, addressline={P.O. Box 49}, city={Budapest}, postcode={1525}, country={Hungary}}

%
%
\begin{abstract}
We aim to advance the state-of-the-art in Quadratic Unconstrained Binary Optimization formulation with a focus on cryptography algorithms. As the minimal QUBO encoding of the linear constraints of optimization problems emerges as the solution of integer linear programming (ILP) problems, by solving special boolean logic formulas (like ANF and DNF) for their integer coefficients it is straightforward to handle any normal form, or any substitution for multi-input AND, OR or XOR operations in a QUBO form. To showcase the efficiency of the proposed approach we considered the most widespread cryptography algorithms including AES-128/192/256, MD5, SHA1 and SHA256.  For each of these, we achieved QUBO instances reduced by thousands of logical variables compared to previously published results, while keeping the QUBO matrix sparse and the magnitude of the coefficients low. In the particular case of AES-256 cryptography function we obtained more than $8\times$ reduction in variable count compared to previous results. The demonstrated reduction in QUBO sizes notably increases the vulnerability of cryptography algorithms against future quantum annealers, capable of embedding around $30$ thousands of logical variables.
\end{abstract}
%
%
%

\end{frontmatter}

\section{Introduction}

The concept of NP-hardness is viewed as the threshold that distinguishes problems that can be solved efficiently with available computational resources from those that are computationally intractable. 
Strategies and heuristics to tackle optimization problems belonging into the NP-hard \cite{10.1145/800157.805047} class have been intensively studies for decades \cite{applegate2011traveling,BLUM1992117,dasgupta2008hardness}. 
During this quest, concepts like Integer Linear Programming (ILP) \cite{GLOVER1986533,tabusearch_glover} 
and Boolean Satisfiability Problem (SAT) \cite{10.1145/800157.805047,doi:10.1126/science.264.5163.1297}
have emerged as fundamental mathematical constructions in formulating real-life problems including number partitioning \cite{Gras2016-st}, Job-Shop Scheduling \cite{venturelli2016quantum}, optimal trading trajectory \cite{10.1145/2830556.2830563}, prime factorization \cite{Dridi2017,jiang2018quantum}, fault detection \cite{Perdomo-Ortiz2015} and Exact Cover \cite{doi:10.1126/science.1057726}.
ILP formulation is also a key concept in resource allocation, data routing and scheduling applied on Field Programmable Gate Arrays (FPGAs) \cite{SUCHA20085770,10.1145/1120725.1121016,5722305} or on Language Processing Units (LPUs) designed for high-throughput and low-latency machine learning applications \cite{9138986,10.1145/3470496.3527405}.
Recently, another formulation of NP-hard problems became popular as well, thanks to the rapid progress shown in the development of quantum computers.
The Quadratic unconstrained binary optimization (QUBO) \cite{Kochenberger2014,Glover2022,glover2019tutorial,Mooney2019MappingNP} is a problem of recent interest due to its relation to quantum annealers \cite{Chancellor2016,McGeoch2014AdiabaticQC,chang2020integer}. 
Unlike classical or gate based quantum computation, the operation of quantum annealers utilise continuous-time Hamiltonian evolution and the adiabatic theorem of quantum mechanics.
In principle, this enables quantum annealers to find the exact global minimum, which is critical when addressing cryptography related problems, as sub-optimal solutions are meaningless in this context. Classical solvers, on the other hand, are not believed to find the optimal solution for these problems.

The common ground of these paradigms is the ability to formulate and solve combinatorial optimization (CO) problems. 
Whichever approach is chosen to address a CO problem, the number of the optimization variables, the density of the encoding and the magnitude of the coefficients are all critical aspects in the solvability of the CO problem \cite{10.1145/3520304.3533952,BIAN2020104609}, especially when talking about quantum devices with limited resources.
In practical quantum annealing, the number of required physical qubits also depends on the mapping onto the topology of the device \cite{Mooney2019MappingNP,BIAN2020104609}.
Therefore, the investigation of efficient encoding strategies has a direct practical application, which was examined for SAT and Hamiltonian Cycle problems\cite{10.1145/3520304.3533952}, Maximum-Weight Independent Set (MIS) \cite{djidjev2018efficient}, Exact Cover \cite{Choi2010AdiabaticQA}, Set Cover \cite{djidjev2023quantum}, Coloring problems \cite{10.3389/fphy.2014.00005} Job-Shop Scheduling \cite{Bianco_Beck_2022} problems and others.

The SAT encoding -- being one of the best known and most fundamental problems in computer science -- has been extensively investigated in the past for many NP-complete problems \cite{10.1145/3520304.3533952}.
We now give a brief overview on phrases and abbreviations used in this work, making easier to follow our reasoning. 

\emph{k-SAT, CNF:} An instance of a \emph{k-SAT} is interpreted as a Boolean formula given in a special form, called \emph{conjunctive normal form} (CNF). 
An example for such a formula is given by $(x \lor y) \land (\overline{x} \lor y)$, where $x$ and $y$ are Boolean variables and $\land$, $\lor$, and $\overline{x}$ are logical operators denoting logical operations \emph{AND}, \emph{OR}, and \emph{NOT}.
Each Boolean variable can be assigned either True or False. Depending on the values assigned to them, the formula as a whole evaluates to either True or False.
We refer to each disjunction (i.e. a logical \emph{OR} operation like $(x \lor y)$) as a clause. 
According to its definition, a \emph{k-CNF} formula is made up from a collection of such clauses, each containing $k$ variables, and joined by conjunctions (i.e. logical \emph{AND} operations).

\emph{DNF, ANF:} Additionaly, a SAT formula is said to be in \emph{disjunctive normal form} (DNF) if it is given by a two-level disjunction-of-conjuctions formula, such as $(x \land y) \lor (\overline{x} \land y)$. 
If \emph{OR} operations are replaced by \emph{XOR} operands than the SAT is said to be in \emph{algebraic normal form}, abbreviated by ANF.

\emph{PES:} Finally, we mention one last form of SAT expressions. 
Similarly to the ANF form, if we incorporate \emph{XOR} operands instead of \emph{OR} operands into the CNF formulation, we end up with a SAT labeled as \emph{product of exclusive sums} (PES).
In this work we show that PES form is a practical extension to CNF, and will turn out to be beneficial to develop a strategy for efficient QUBO encoding.

\emph{min-term}, $\mathbf{x}_{\textrm{min}}$: A SAT formula is satisfied when it evaluates to \emph{True}, and each corresponding set of boolean variables is called a \emph{min-term} and labeled by $\mathbf{x}_{\textrm{min}}$.

Special SAT forms were also exploited in the formulation to solve security critical problems including cryptography and hash algorithms such as Advanced Encryption Standard (AES), MD5, Secure Hash Algorithm (SHA) algorithms \cite{otpuschennikov2016encoding,9690063,a15020033,iet-ifs.2017.0176}, Logical Cryptanalysis \cite{Massacci2000}, or data encryption algorithms\cite{Lafitte2014}. 
In general, we can consider an encoding protocol to be compromised, if one can construct an inverse for the function $f$ implementing the algorithm of the given cryptography or hash protocol.
In the SAT formulation the inversion problem of $f$ is challenged by reducing it to the problem of finding a satisfying assignment of SAT instances in a CNF form, as it follows from the Cook-Levin theorem \cite{Cook1971TheCO}. 
Therefore, many existing methods for generating cryptography SAT instances use essentially the Tseitin\cite{Tseitin1983OnTC} transformations \cite{10.1007/11814948_13,cryptoeprint:2014/239,10.1007/978-3-540-72788-0_36}. 
Existing cryptography-focused solvers such as \emph{CryptoSAT} \cite{iet-ifs.2017.0176} and \emph{cryptominisat} \cite{DBLP:conf/sat/SoosNC09} specifically support \emph{CNF} based SAT encoding of cryptography problems. Additionally, \emph{CryptoSAT} also supports PES and can work in conjunction with Bosphorus which aims to simplify \emph{CNF} and \emph{ANF} formulae \cite{8715061}.  This is especially useful in cryptography applications, where operations in Galois Fields (or finite fields) with $2^k$ elements are commonplace ($k$ is an integer defining the size of the finite field).

As SAT--to--QUBO encoding strategies turned out to have a significant impact on the solution quality of quantum annealing \cite{Kurin2019ImprovingSS,10.1145/3583133.3596330}, numerous research work have been conducting on QUBO transformation methods \cite{doi:10.1142/9789812778215_0008,VERMA2022100594,BIAN2020104609,verma2021variable,Chancellor2016,Choi2010AdiabaticQA,10.1145/3583133.3596394,10.1007/978-3-031-36030-5_3,wolf2023automatic,Karimi2017PracticalIM,djidjev2023quantum}. 
According to the study presented in Ref.~\cite{BIAN2020104609} the SAT--to--QUBO transform can be formulated as a problem in Satisfiability or Optimization Modulo Theories (SMT/OMT) \cite{10.1007/978-3-642-12002-2_8,10.1145/2699915} on the theory of linear rational arithmetic.
By encoding the whole SAT problem in one step would typically require the introduction of many additional ancillary boolean variables. These extra variables would result in very large SMT/OMT formulas: solving the resulting problem would become possibly even harder than the original SAT problem.
In order to overcome this issue scalable “divide-and-conquer” approach was proposed for the SAT--to--QUBO transformation.
In this regard, the special forms of SAT instances mentioned above can be used to engineer encoding strategies to reduce the variable count in the formulation of CO problems \cite{Kucera2017ALB}.
The work of Ref. \cite{10.1145/3520304.3533952}, for example, presented two novel QUBO formulations for $k$-SAT and Hamiltonian Cycles that scale significantly better than preceding approaches. For $k$-SAT the growth of the QUBO matrix was reduced from $\mathcal{O}(k)$ to $\mathcal{O}(\textrm{log} (k))$, while for Hamiltonian Cycles the new approach gives a QUBO matrix scaling linearly in the number of edges and logarithmically in the number of nodes.
The reported improvement was achieved by an iterative squaring technique for the CNF form of SAT.
Additionally, a systematic meta-algorithm was introduced in the study of Ref.~\cite{electronics12163492}, facilitating the design of automatized QUBO encoding of $3$-SAT problems.
The authors of this work pointed out the importance of patterns in SAT formulations and use their properties in algorithmic methods to encode QUBO instances. 
The study in Ref.~\cite{Verma2021VariableRF} introduced a logarithmicaly scaling method to encode equality and inequality constraints that was further enhanced by research of Ref.~\cite{10.1145/3631908.3631929} providing an asymptotic improvement in the number of variables.

\emph{Contribution of the current work:} In this work we further extend the QUBO encoding introduced in the work of Ref.~\cite{10.1145/3631908.3631929} by the inclusion of parity (i.e.~\emph{XOR}) formulae. 
This extension turned to be a foundation in designing an efficient strategy to encode various SAT normal forms, applicable to formulate QUBO instances for complex mathematical designs as building blocks.
As SAT instances, in general, contain higher than quadratic-order clauses, the QUBO transform of a SAT will necessarily involve substitution variables. 
The aim of this work is to find a SAT--to--QUBO mapping implying the fewest substitution variables as possible.
We demonstrate the improved encoding strategy on cryptography constructions, such as the AES-128/192/256 encoding schemes and the MD5, SHA1 and SHA-256 hash functions.
These applications proved the saving of thousands of QUBO variables when encoding them.

The rest of the paper is organized as follows: in Sec.~\ref{sec:exhaustive_search} we describe the core component of our encoding methodology, relying on the ILP formulation of the SAT--to--QUBO mapping.
While the ILP methodology is limited to small sized problems, it enabled the discovery of generic patterns to scale-up the QUBO encoding of SAT normal forms. We discuss these patterns in Sec.~\ref{eq:QUBO_normal_forms}.
We apply our method to construct the QUBO representation of cryptography constructions and discuss our results in Sec.~\ref{sec:crypto}.
Finally, we conclude our work in Sec.~\ref{sec:conclusions}.













\section{The theory of finding QUBO Encoding Patterns} \label{sec:exhaustive_search}

A QUBO problem with $N$ binary variables is represented by a minimization (or maximization) of 
\begin{equation}
    \sum\limits_{i, j=0}^{N-1} x_i Q_{i,j} x_j\;, 
\end{equation}
 where $x\in \{0,1\}^N$ and $Q$ is a symmetric matrix from $\mathbb{R}^{N\times N}$.
Our method to find a suitable and highly compressed SAT--to--QUBO mapping to encode the addressed optimization problem relies on solving an associated \emph{ILP} problem.
At this point it is important to mention, that our objective is to find reusable patterns for SAT--to--QUBO mappings and not to solve the SAT problem via QUBO formulation.
Therefore, to find appropriate encoding patterns we consider tractable (i.e. small-scaled) SAT instances for which we can find all the \emph{min-terms}.
In general, an ILP seeks to minimize an objective 
\begin{equation}
\min\limits_{x} (c^\intercal x)
\end{equation}
subjected to constraints
\begin{equation}
\mathbf{A_{ub}}\mathbf{x}\le \mathbf{b_{ub}}\;,\quad \mathbf{A_{eq}}\mathbf{x}=\mathbf{b_{eq}}\;,\quad \mathbf{l}\le \mathbf{x}\le \mathbf{u}\;,    \label{eq:general_ILP_constraints}
\end{equation}
where the real vectors $\mathbf{c}$, $\mathbf{l}$, $\mathbf{u}$ are taken from $\mathbb{R}^N$.
The number of elements in the vectors $\mathbf{b_{ub}}$ and $\mathbf{b_{eq}}$ refer to the number of inequality and equality constraints, respectively. Matrices $\mathbf{A_{ub}}$ and $\mathbf{A_{eq}}$ have rows determined by the number of constraints and columns equal to the involved elements in vector $\mathbf{x}$.
A potential solution vector $\mathbf{x}$ is said to be feasible, if all of the constraints are satisfied.  
If there does not exist any $\mathbf{x}$ vector which is feasible, then the ILP system is said to be infeasible.
In this section we describe the formulation of the \emph{ILP} problem to find the optimal SAT--to--QUBO mapping. 
Firstly, we introduce a set of binary substitution variables $\mathbf{s}=\{s_i\}$, with $i\in\{0,1,\dots,N_s-1\}$, $s_i\in\{0,1\}$ and the number of new variables $N_s$ specified later.
In the literature, often the more general term of ancillary variables is used, but we prefer the term substitution as it emphasizes the importance that the substitution vriable can be isolated as an equivalence relationship by boolean formulae.

In principle, the valid values of the variables $\mathbf{s}$ are determined via a problem specific binary-input function $\mathbf{s}_{\textrm{valid}}=\mathbf{s}_{\textrm{valid}}(\mathbf{x})$.
The related literature provides countless examples to construct the function $\mathbf{s}_{\textrm{valid}}(\mathbf{x})$ \cite{Peng2019,mandal2020compressed,Domino2022}.
Usually these examples involve the reduction of higher order terms, such as the transform of a 3-SAT term $x_1 \land x_2 \land x_3$ into QUBO by introducing a substitution variable $s=s_{\textrm{valid}}(\mathbf{x}) = x_1\cdot x_2$ and append a penalty term to the QUBO formula giving a zero energy only for $s=s_{\textrm{valid}}(\mathbf{x})$ and is greater than zero otherwise. 
For this specific case Rosenberg showed that $g(\mathbf{x}, s)=3s + x_1 x_2 - 2x_1 s - 2x_2 s$ is a well chosen function to fulfill these requirements \cite{Rosenberg}. However, for larger problems the outlined strategy to define substitution variables turns to be wasteful in the variable count as one can find better encoding by considering larger portions of the SAT formula and finding nontrivial substitutions and penalty terms as it was demonstarted in the work of Ref.~\cite{10.21203/rs.3.rs-3471221/v1}.
Here we describe our general strategy of finding compact encoding QUBO formulae. 
To this end we introduce a quadratic formula $g(\mathbf{x},\mathbf{s})$ as a function of binary substitution variables encompassed by the vector $\mathbf{s}$.
We expect the quadratic function $g(\mathbf{x},\mathbf{s})$ to take on a return value $0$ for any min-term input $\mathbf{x}_{\textrm{min}}$ and for the substitution variables $\mathbf{s}_{\textrm{valid}}(\mathbf{x}_{\textrm{min}})$:
\begin{equation}
    g\bigg(\mathbf{x}_{\textrm{min}}, \mathbf{s}=\mathbf{s}_{\textrm{valid}}(\mathbf{x}_{\textrm{min}})\bigg)=0\;.\label{eq:g_condition}
\end{equation}
For any other binary inputs $g(\mathbf{x},\mathbf{s})$ should give values larger than $0$:
\begin{equation}
    g\bigg(\mathbf{x}, \mathbf{s}\bigg)>0\;, \label{eq:g_condition2}\qquad \textrm{if: } \mathbf{x}\neq\mathbf{x}_{\textrm{min}} \;\textrm{or}\; \mathbf{s}\neq\mathbf{s}_{\textrm{valid}}(\mathbf{x}_{\textrm{min}})\;.
\end{equation}
A quadratic function $g(\mathbf{x},\mathbf{s})$ fulfilling these requirements will encode the same optimization problem as the original SAT instance.
The function $g(\mathbf{x},\mathbf{s})$ can be given in a general form
\begin{equation}
\begin{split}
    g(\mathbf{x},\mathbf{s}) = \sum\limits_{i=0}^{N-1} c_i x_i + \sum\limits_{i,j=0}^{N-1} d_{i,j}x_i x_j +  \\
    \sum\limits_{i=0}^{N-1}\sum\limits_{i=0}^{N_s-1} e_{i,j}x_i s_j + \sum\limits_{i=0}^{N_s-1} f_{i}s_i + \sum\limits_{i,j=0}^{N_s-1} g_{i,j}s_i s_j + h\;,
\end{split}    \label{eq:general_g}
\end{equation}
with unknown integer coefficients $\mathbf{c}=\{c_i\}$, $\mathbf{d}=\{d_{i,j}\}$, $\mathbf{e}=\{e_{i,j}\}$, $\mathbf{f}=\{f_i\}$, $\mathbf{g}=\{g_{i,j}\}$ and an integer constant $h$.
Finding the coefficients $\mathbf{c}$, $\mathbf{d}$, $\mathbf{e}$, $\mathbf{f}$, $\mathbf{g}$ and $h$ satisfying Eqs. (\ref{eq:g_condition}) and (\ref{eq:g_condition2}) is the goal of an \emph{ILP} optimization problem.
Furthermore, the solution of the \emph{ILP} problem should also determine an un-ambigous $\mathbf{s}_{\textrm{valid}}(\mathbf{x}_{\textrm{min}})$ function, so the search is also extended over the variables $\mathbf{s}$, while we assume all the \textrm{min-terms} of the original SAT instance to be known.
This \emph{ILP} system, however, suffers from higher order dependencies across the equation system as we have cubic and quartic terms of unknown variables in Eq.~(\ref{eq:general_g}), making it challenging for state-of-the-art solvers to make progress even for problems involving a small number ($\sim10$) of unknown variables.
A simplified \emph{ILP} system can be constructed to increase the chance of success when searching for two or more substitution variable functions.
We formulate the corresponding \emph{ILP} problem by performing an exhaustive search over all the possible element-variations of $\mathbf{x}$ and $\mathbf{s}$ (thus, variables $\mathbf{x}$ and $\mathbf{s}$ are not treated as unknowns) and find the coefficients in Eq.~(\ref{eq:general_g}) fulfilling the constraints (\ref{eq:g_condition}) and (\ref{eq:g_condition2}). 
Inequality constraints of Eq. (\ref{eq:g_condition2}) can be turned into equality constraint by the introduction of binary slack variables $k_i$ and integer variables $t_i\in\{1,\dots, t_i^{\textrm{max}}\}$ with $i\in\{0, \dots, 2^{N_s}-1\}$ by insisting \cite{a15020033}
\begin{equation}
    \tilde{g}(\mathbf{x}, \mathbf{s}_i, k_i, t_i) = g(\mathbf{x}, \mathbf{s}_i)-k_i t_i=0\;, \label{eq:g_condition3}
\end{equation}
where $\mathbf{s}_i$ is the $i$-th bit-string of the substitution variables during the exhaustive search, such that $\mathbf{s}_i$ is nothing else than the binary representation of the integer $i$.
Then, the \emph{ILP} problem can be formulated as follows
\begin{equation}
\mathcal{P} = \bigg(\mathcal{P}_{\textrm{min-term}}\quad \wedge \bigg(\bigwedge\limits_{\mathbf{x}\neq\mathbf{x}_{\textrm{min}}}\mathcal{P}_{\textrm{non-min-term}}(\mathbf{x})\bigg)\bigg) = 1 , \label{eq:ILP-full}
\end{equation}
where
\begin{equation}
\begin{split}
    \mathcal{P}_{\textrm{min-term}} &= \\
    g(\mathbf{x}_{\textrm{min}}, \mathbf{s}_0)&=0 \quad \vee \quad  g(\mathbf{x}_{\textrm{min}}, \mathbf{s}_1)=0 \quad \vee \quad \dots  \\
    &\dots \quad\vee \quad g(\mathbf{x}_{\textrm{min}}, \mathbf{s}_{2^{N_s}-1})=0 \;, 
\end{split}\label{eq:ILP}
\end{equation}
and 
\begin{equation}
\begin{split}
    \mathcal{P}_{\textrm{non-min-term}}(\mathbf{x}) &= \\
    \tilde{g}(\mathbf{x}, \mathbf{s}_0, k_0=1)&=0 \quad \wedge \quad  \tilde{g}(\mathbf{x}, \mathbf{s}_1, k_1=1)=0 \quad \wedge  \\
    &\dots \quad\wedge \quad \tilde{g}(\mathbf{x}, \mathbf{s}_{2^{N_s}-1}, k_{2^{N_s}-1}=1)=0 \;.
\end{split}\label{eq:ILP_non_minterm}
\end{equation}
In this context the goal of the \emph{ILP} problem is to find a feasible solution to the constraint (\ref{eq:ILP-full}) (without any objective function to be minimized).
Eq.(\ref{eq:ILP}) can be translated into ILP compatible format by:
\begin{equation}
\begin{split}
    \mathcal{P}_{\textrm{min-term}} &= \\
    \tilde{g}(\mathbf{x}_{\textrm{min}}, \mathbf{s}_0)&=0 \quad \wedge \quad  \tilde{g}(\mathbf{x}_{\textrm{min}}, \mathbf{s}_1)=0 \quad \wedge \quad \dots  \\
    &\dots \quad\wedge \quad \tilde{g}(\mathbf{x}_{\textrm{min}}, \mathbf{s}_{2^{N_s}-1})=0 \;, 
\end{split},\label{eq:ILP_gtilde}
\end{equation}
where we have replaced the polynomial $g$ with $\tilde{g}$ and used $\wedge$ clauses instead of $\vee$ operators. 
This transformation is justified as valid substitution variable maps are determined by clauses being satisfied with $k_i=0$. 
The conjunctive clauses in (\ref{eq:ILP_non_minterm}) and (\ref{eq:ILP_gtilde}) are straightforward to convert into ILP as all of the involved clauses need to be satisfied simultaneously, resulting in a set of linear constraints of the form of Eq.~(\ref{eq:general_ILP_constraints}) with respect to the unknown coefficients in the polynomial $g$.
Regarding the constructed ILP, the condition (\ref{eq:g_condition3}) ensures two important properties.
Firstly, as the slack variables $t_i$ are greater than or equal to $1$ and $k_i$ is either $0$ or $1$,  constraint (\ref{eq:g_condition2}) will be satisfied when the constructed \emph{ILP} is solved.
Secondly, by summing up the values of $k_i$ in (\ref{eq:ILP_gtilde}) we can deduce the number of bit-strings $\mathbf{s}_i$ satisfying Eq.~(\ref{eq:g_condition}).
If $\sum_i k_i = 2^{N_s}$ holds on, then the \emph{ILP} has not found any valid substitution variable vector satisfying (\ref{eq:g_condition}), and we need to increase $N_s$.
When $\sum_i k_i = 2^{N_s}-1$, then there is only a single bit-string $\mathbf{s}$ satisfying (\ref{eq:g_condition}) and the number of the substitution variables should be considered optimal.
Finally, we can assume that the number of the substitution variables can be decreased then $\sum_i k_i < 2^{N_s}-1$.
The equation $\sum_i k_i = 2^{N_s}-1$ implicitly determines the optimal number $N_s$ of the substitution variables.
The solution of the \emph{ILP} problem implicitly also determines the unique substitution variable vector $\mathbf{s}=\mathbf{s}_{\textrm{valid}}(\mathbf{x}_{\textrm{min}})=\mathbf{s}_i$ corresponding to $k_i=0$ in Eq~(\ref{eq:g_condition3}).

The procedure described above can be generalized to a situation when the SAT instance can be satisfied with multiple \emph{min-terms}.
This is especially relevant in cases when the original SAT formula is partitioned into smaller parts and each part is individually encoded into a QUBO. (The resulting quadratic functions then can be summed up into a single unified QUBO.)
In this case the individual SAT parts might have multiple \emph{min-terms}, including the one satisfying the whole SAT formula. 
Equation (\ref{eq:ILP_gtilde}) is then modified by the conjuction of multiple \emph{min-term} disjunctions resembling the \emph{CNF} form of a SAT:
\begin{equation}
\begin{split}
    \mathcal{P}_{\textrm{multi-min-terms}} = \\
    \bigg(\tilde{g}(\mathbf{x}_{\textrm{min}}^{(1)}, \mathbf{s}_0)=0 \quad\wedge\quad  &\dots \quad \wedge \quad  \tilde{g}(\mathbf{x}_{\textrm{min}}^{(1)}, \mathbf{s}_{2^{N_s}-1})=0\bigg) \\
       &\wedge \\
        \bigg(\tilde{g}(\mathbf{x}_{\textrm{min}}^{(2)}, \mathbf{s}_0)=0 \quad\wedge\quad  &\dots \quad\wedge \quad  \tilde{g}(\mathbf{x}_{\textrm{min}}^{(2)}, \mathbf{s}_{2^{N_s}-1})=0\bigg) \\
        &\wedge \\ \dots
\end{split} \label{eq:ILP2}
\end{equation}
Equation (\ref{eq:ILP2}) is than inserted into Eq.~(\ref{eq:ILP-full}) instead of $\mathcal{P}_{\textrm{min-term}}$ and solved for the coefficients in Eq.~(\ref{eq:general_g}).
By solving this \emph{ILP} we retrieve a unique and valid substitution bit-string $\mathbf{s}$ for each min-term $\mathbf{x}_{\textrm{min}}^{(l)}$.
Additionally, at the minimal number of substitution variables the count of $k_i=0$ variables is equal to the number of the \emph{min-terms}.

When comes to solve the \emph{ILP} problems for the coefficients, the so-called symmetry breaking approach \cite{soeken2020determining} can be exploited to reduce the search space.
The optimization problem posses a symmetry if inter-changing at least two optimization variables in the solution we still end up with a valid solution.
The main idea of symmetry breaking is to impose extra constraints on the possible solutions of the optimization problem, like imposing non-strict inequality constraints on the variables.

In relation with the outlined strategy we should mention another subtle detail.
We classify QUBO substitution variables into three categories: (i) pre-constrained (ii) strong and (iii) weak.  
A \emph{pre-constrained} substitution is represented by a variable whose constraints have already been encoded into a QUBO form.
Such a situation occurs when we have already encoded a substitution variable in a partition of the SAT, and would like to reuse the same substitution in other partitions again to encode identical variable patterns. 
This way the constraint on the valid value of the substitution variables is guaranteed, as at the end all SAT partitions are going to be summed up into a single QUBO. 
A \emph{strong substitution} function is one where the bad substitution values 
always give a higher value of $g(\mathbf{x}, \mathbf{s})$ than the correct intended substitution values.  
Therefore, the substitution function is exactly represented as the encoding makes the minimum only occur at correct substitution values.  A \emph{weak substitution} function is one where the substitution values are ambiguous, i.e. different substitution variables can be used to get the minimum of $g(\mathbf{x}, \mathbf{s})$.  
This distinction is noteworthy as re-use of substitution variables may in some cases be possible.  Pre-constrained substitution variables significantly reduce the number of constraints in the ILP.  Strong substitution variables increase the number of constraints, but can be reliably re-used.

To solve the ILP problem, we can utilize a counter-example guided optimization strategy which can play an important role on difficult problems.  In this strategy, we add all of the constraints representing non-\emph{min-terms} (i.e. Eq.~(\ref{eq:ILP_non_minterm})), and then add the \emph{min-term} constraints (i.e. Eqs. from (\ref{eq:ILP2})) one by one.  If no solution is found, it becomes clear that more substitution variables are needed.  Otherwise, the coefficients can be checked against for all the remaining \emph{min-term} constraints, and the first one which does not satisfy equality to zero with all substitution values is then added to the next ILP call.  This is particular beneficial in incremental ILP solvers that maintain their internal solver state.

\section{Obtained QUBO Patterns for Normal Forms} \label{eq:QUBO_normal_forms}

As the solution of the general QUBO encoding procedure described in the previous section is extremely challenging for large problems, deriving encoding patterns for special SAT forms is of high importance.
To fulfill this goal in this work we generalize our previous result reported in \cite{10.1145/3631908.3631929}, which was initially developed for the population counting set function, enabling ones to encode constraints like $L\leq\sum x_i\leq H$.
The generalized approach is applicable to numerous function types, including multi-input conjunctions and parity functions (expressed via multi-input \emph{XOR} clauses) as found in DNF and ANF forms, respectively.
Here we summarize our numerical results obtained by solving Eq.~(\ref{eq:ILP-full}) and in a later section we will examine and justify the discovered patterns.

\subsection{Parity encoding: multi-input XOR clauses} \label{sec:XORencoding}

Let us first examine parity encoding. 
The very first thing to decide is the way the substitution variables are represented.
The idea here is that the parity function decides on a sum of input variables whether it is odd (return value $1$) or not (return value  $0$).
To turn this function into a QUBO form (supposing we limit the input variables to binary ones) we search for the minimum of
\begin{equation}
    \left(\sum\limits_i x_i - T\right)^2\;, \label{eq:parity_QUBO}
\end{equation}
where the substitution variable $T$ can take on values from the set of odd integers greater than $0$.
If the sum of the variables is odd, one can find a suitable value for $T$ to cancel the sum ending in a minimal value of $0$ for the quadratic expression, indicating that the sum is odd.
Otherwise, the substitution variable $T$ can not cancel the sum, giving nonzero output for the quadratic expression even at its minimum.
We now look for the optimal binary representation of $T$.
Odd integers can be given in a form $T=2t-1$ with $1\leq t\leq t_{max}$. 
$t_{max}$ is determined according to the maximal value of $\sum_i x_i$ that $T$ needs to cancel. 
In case we have $N$ binary variables in the sum, we get $t_{max}=\lceil N/2 \rceil$. 
Then, if $t_{\max}$ is equal to a power of $2$, the most optimal encoding of $t$ is simply taking its binary expansion:
\begin{equation}
    t = 1 + \sum\limits_{i=0}^{N_{subs}-1} 2^{i} s_i \label{eq:t}
\end{equation}
with a number of substitution variables $N_{subs}= \log_2 (t_{max})$.
In a more general case when $t_{max}$ is not a power of $2$ we propose a modified expression to encode a slack variable covering the needed range:
\begin{equation}
    t = 1 + \left(t_{max} - 1 -\sum\limits_{i=0}^{N_{subs}-2} 2^{i}\right) s_{N_{subs}-1} + \sum\limits_{i=0}^{N_{subs}-2} 2^{i} s_i\;, \label{eq:relaxed_binary}
\end{equation}
where $N_{subs}=\lceil\log_2 (t_{max})\rceil$. 
In this expression the most significant binary variable $s_{N_{subs}-1}$ has a coefficient whose value is relaxed compared to a conventional binary number representation in front of $s_{N_{subs}-1}$. The relaxed binary expansion is general, can be used either when $t_{max}$ is a power of $2$ or not.
With this modified binary expansion form, however, there occurs no overflows in the covered range determined by $[1,t_{\max}]$.

By solving the ILP problem given by Eq.~(\ref{eq:ILP-full}) one can prove the optimal number of substitution variable. For smaller sized problems we have found the encoding of the multi-input \emph{XOR} function given by (\ref{eq:parity_QUBO}) and \ref{eq:t}) to be optimal in the number of binary substitution variables.

\subsection{Range constraints: multi-input OR clauses} \label{sec:ORencoding}

Surprisingly, the key result of our work revealed that for other binary functions (such as multi-input \emph{OR} or population count) the number of substitution variables can be decreased with a non-trivial encoding pattern.
This is achieved via a form 
\begin{equation}
    \left(\sum\limits_i x_i - (2t - 1)\right)\cdot\left(\sum\limits_i x_i - (2t-2)\right)\;, \label{eq:multi_input_QUBO}
\end{equation}
instead of direct squaring done in Eq.~(\ref{eq:parity_QUBO}).
We explain this function as follows. 
A multi-input \emph{OR} function on variables $\{x_0,x_1,\dots, x_{N-1}\}$ and given by
\begin{equation}
    \bigvee\limits_{i=0}^{N-1} x_i \label{eq:multiOR}
\end{equation}
can be expressed in the form
\begin{equation}
    \sum\limits_{i=0}^{N-1} x_i - T = 0\;,\label{eq:encoded_multiOR}
\end{equation}
where a suitable substitution variable $1\leq T\leq N$ can be found for the equality.  
In case having such $T$, the multi-input \emph{OR} function (\ref{eq:multiOR}) can be satisfied to give \emph{True} output.
Otherwise, Eq.(\ref{eq:multiOR}) is giving output \emph{False} and (\ref{eq:encoded_multiOR}) gets violated.
As the binary and integer variables can change only by $1$, (\ref{eq:multi_input_QUBO}) can take on only non-negative values, while giving $0$ only when fulfilling (\ref{eq:encoded_multiOR}) with $T=2t-1$ or $T=2t-2$.
Therefore, the substitution variable $t$ is required to span a shorter range of $1\leq t\leq t_{max}$, where $t_{max}=\lceil (N+1)/2 \rceil$.
The shorter range requires one less bit in the binary expansion of $t$ resulting in saving one bit (i.e. one QUBO variable) in the encoding when using (\ref{eq:multi_input_QUBO}) instead of \ref{eq:parity_QUBO}).
For the binary expansion of $t$ we can use (\ref{eq:relaxed_binary}) to cover the range $1\leq t\leq t_{max}$.
An important point here is to understand the mechanism of (\ref{eq:multi_input_QUBO}) on saving one bit in the encoding of the substitution variable. 
As one can see, a single combination of substitution variables in this formula is used to encode two different values at once, which are $2t-1$ and $2t-2$ ($t$ is expressed in terms of substitution variables $s_i$ via a reduced binary expansion of Eq.~(\ref{eq:relaxed_binary})).
This dualism is the key reason for the saving of one bit in the encoding.
In the case of even $N$ a subtle detail to notice is that $2t_{max}-1=N+1$, seemingly overflowing the range limit $N$. 
However, the left side of Eq.~(\ref{eq:multi_input_QUBO}) will becomes zero for $t=t_{max}$, thus the QUBO encoding (\ref{eq:multi_input_QUBO}) will not take on negative values or any false minima.

At this point we also need to mention that the relaxed binary encoding of (\ref{eq:relaxed_binary}) brings in an ambiguity into the encoding, as more than one combinations of the binary variables $\{s_i\}$ might give the same substitution variable $t$.
Additionally, the binary expansion (\ref{eq:relaxed_binary}) can be tweaked by modifying the remaining coefficients besides the most significant one in front of $s_{N_{subs}-1}$, giving further possible encoding possibilities to cover the set $\{1,2,3,\dots,N\}$.
Having suitable $N_{subs}$-variable functions $f_1$ and $f_2$ we can generalize (\ref{eq:multi_input_QUBO}) into the form
\begin{equation}
    \left(\sum\limits_i x_i - f_1(\{s_i\})\right)\cdot\left(\sum\limits_i x_i - f_2(\{s_i\})\right)\;, \label{eq:general_multi_input_QUBO}
\end{equation}
where functions $f_1(\{s_i\})$ and $f_2(\{s_i\})$ must cover all elements in the set $\{1,2,3,\dots,N\}$.
When $t_{max}$ is not a power of $2$, the encoding of the combinations of the binary substitution variables $\{s_i\}$ is larger than the range needed to be spanned.
Therefore, $f_1$ and $f_2$ don't need to give a unique cover of the spanned range, they might coincide for some inputs.  

Equation (\ref{eq:general_multi_input_QUBO}) is the most general form to encode the multi-input \emph{OR} formula. The terms on the right and left must differ precisely by $1$ (or equal to each other), otherwise the product might give negative values for some inputs $\{x_i\}$ obscuring the QUBO encoding by providing incorrect minima.
A simple example for $f_1(\{s_i\})$ and $f_2(\{s_i\})$ can be given for the case of a multi-input \emph{OR} formula with $N=10$ variables $x_i$. In this case $t_{max}=6$, thus $N_s=3$ and 
\begin{equation}
    f_1(s_1,s_2,s_3) = 5s_3 + 3s_2 + 2s_1
\end{equation}
with
\begin{equation}
    f_2(s_1,s_2,s_3) = 5s_3 + 2s_2 + s_1 + 1
\end{equation}
become an alternative choice to encode the substitution variables besides (\ref{eq:multi_input_QUBO}).
Practically speaking, the finding of suitable encoding functions translates onto the covering set problem 
paving the way for deeper insight into fostering optimal QUBO encoding strategies.
By solving the ILP problem given by Eq.~(\ref{eq:ILP-full}) one can numerically prove that no other encoding can be found encompassing less substitution variables.

Besides multi-input \emph{OR} clauses the outlined encoding strategy is also applicable to encode other multi-input binary functions, including population count constraints limiting the sum of variables $\{x_i\}$ into a specific range via
\begin{equation}
    L \leq \left(\sum\limits_{i=0}^{N-1} x_i\right) \leq H\;. \label{eq:range_constarint}
\end{equation}
Range constraints expressed with strict inequalities (i.e. $<$ and $>$) can be transformed into the form (\ref{eq:range_constarint}) by subtracting $1$ from $H$ or adding $1$ to $L$.
In this case the range $T$ to be spanned by substitution variables is determined by $T=H-L+1$ (the lowest and highest values are also included in the range).
As the population count $\sum x_i$ can exceed $H$ or can be less than $L$, extra caution needs to be payed preventing from overstepping the boundaries $L$ and $H$ by the minima of the QUBO encoded form of the range constraint, otherwise it will convey false minima.
For a range $T$ (being either even or odd) the function $f_1$ can be encoded as
\begin{equation}
    f_1(\{s_i\}) = L + \left(H - L -\sum\limits_{i=0}^{N_{subs}-2} 2^{i+1}\right) s_{N_{subs}-1} + \sum\limits_{i=0}^{N_{subs}-2} 2^{i+1} s_i\;, \label{eq:f1_general}
\end{equation}
with $N_{subs} = \lfloor\log_2 (T)\rfloor$ and  similarly to Eq.~(\ref{eq:relaxed_binary}), a relaxed coefficient in front of the most significant bit $s_{N_{subs}-1}$.
The distance between any pair of nearest values of $f_1$ -- ensured by the least significant bits -- never exceeds $2$.
Some $\{s_i\}$ combinations can coincide in the resulting value of $f_1$ (for even $T$) or differ only by $1$ (for odd $T$).
Therefore, a suitable $f_2$ can be constructed by simply offsetting $f_1$ by one:
\begin{equation}
    f_2(\{s_i\}) = f_1(\{s_i\}) + 1\;. \label{eq:f2_general}
\end{equation}
Functions $f_1$ and $f_2$ cover the whole range of $\{L, L+1, \dots, H\}$ with alternating values coming from $f_1$ and $f_2$.
An important detail here is that when $f_1=H$ then $f_2$ will take on the value of $H+1$, overstepping the given range.
However, similarly to the QUBO encoding (\ref{eq:multi_input_QUBO}), the left side of Eq.~(\ref{eq:general_multi_input_QUBO}) will prevent from picking up negative values (or false minima) by
the QUBO objective function.
Further valid $f_1$ and $f_2$ encodings can be generated when the relaxation of the coefficients is applied to the other coefficients as well, provided that the difference between any pair of nearest elements among the possible values of $f_1$ will not exceed $2$.
Because of this reason, the relaxation of the least significant bit is not allowed.
The rule of thumb of the relaxation is that the coefficient in front of any $s_i$ must be less or equal to the sum of the coefficients in front of $s_j$-s with $1\leq j<i$ or larger than this sum at most by $2$, provided that the coefficients are in ascending order from $i=0$ to $i=N_{subs}-1$.
The amount of the relaxation, i.e. the accumulated decrease of the coefficients, equals to $2^{N_{subs}+1}-T$, as $2^{N_{subs}+1}$ is the maximum of the range that can be spanned with $N_{subs}$ binary variables.
Following this rule and Eq.~(\ref{eq:f2_general}) one can construct a plenty of valid encodings $f_1$ and $f_2$.

By summarizing our findings we formulate the Root Squeezing Theorem and the Range Encoding Theorem which states that all range constraint QUBOs are capable of being expressed with minimal ancillary variables via a variant of the form given by Eq.~(\ref{eq:general_multi_input_QUBO}).
\newtheorem{theorem}{Theorem}
\begin{theorem}[Root Squeezing Theorem]
Given two integer linear equations $g(X), h(X) \in \mathbb{Z}$ both ascending or descending when their input is increased, and whose roots can be found at $q_0, q_1 \in \mathbb{Z}$, respectively. 
If $|q_0-q_1|\le 1$ then $g(X)h(X)\ge 0 \;\forall X \in \{0,1\}^n$ and  $g(X)h(X) = 0$ only when $X=q_1$ or $X=q_2$.
\end{theorem}
The proof of the theorem is straightforward.
In the context of our example of Eq.~(\ref{eq:general_multi_input_QUBO}) $g=\sum_i x_i - f_1(\{s_i\})$ and $h=\sum_i x_i - f_2(\{s_i\})$ for a specific configuration of the $\{s_i\}$ substitution variables. 
The condition of the roots $|q_0-q_1|\le 1$ is ensured by the functions $f_1$ and $f_2$ for all configurations of the substitution variables. 
\begin{theorem}[Range Encoding Theorem]
Given two natural \\ numbers $L$ and $H$ such that $L<H$, and $N\geq H$ binary variables $x_i\in\{0,1\}$ ($i\in\{0,\dots, N-1\}$).
An upper bound on the number of the required substitution variables $N_{subs}$ to encode the range constraint $L\leq x\leq H$ into a QUBO form (with the population count $x= x_0 + x_1 + \dots + x_{N-1}$), is given by $N_{subs} = \lceil\log_2 (H-L+1)\rceil - 1$.
\end{theorem}
The theorem is a direct consequence of the Root Squeezing Theorem, as by constructing a suitable, $\{s_i\}$ dependent functions $f_1$ and $f_2$ -- given for example by Eqs.~(\ref{eq:f1_general}) and (\ref{eq:f2_general}) -- one can fulfill the criteria of the Root Squeezing Theorem by formulating the appropriate functions $g(X)=\sum_i x_i - f_1(\{s_i\})$ and $h(X)=\sum_i x_i - f_2(\{s_i\})$.
The product $g(X)h(X)$ gives the QUBO encoding of the range constraint with the number of substitution variables according to the Range Encoding Theorem.
As such an encoding always exists, the theorem gives an upper bound on the number of the required substitution variables that are always sufficient to encode the range constraint.
While we believe that this upper bound also equals to the lower bound, we could prove the optimal count of the substitution variables only on small sized problems via exhaustive search by solving the associated ILP problem described in Sec.~\ref{sec:exhaustive_search}.

\subsection{QUBO encoding of \emph{CNF}, \emph{DNF} and \emph{ANF} forms} \label{sec:subst_encoding}

In many applications the result of SAT clauses, as intermediate results, are further processed in the computational procedure. 
This is happening also in the \emph{CNF}, \emph{DNF} and \emph{ANF} SAT forms where multi-input binary terms (clauses) are processed as inputs for further conjunctive or disjunctive binary operations.
Here we outline a method to accomplish the same effect in the QUBO encoded form by assigning an extra binary variable $r$ to the result of the evaluated SAT sub-clauses. These extra variables can be reused in further QUBO encodings, enabling to merge all of the SAT sub-clauses into a single unified QUBO instance.  
In the case of a \emph{XOR} (labeled by $\oplus$) clause $r$ is defined via a strong equivalence
\begin{equation}
r\leftrightarrow\bigoplus\limits_{i=0}^{N-1} x_i \;. \label{eq:XOR_strong_equivalence}
\end{equation}
$r=1$ when the multi-variable $XOR$ clause is satisfied, and $r=0$ otherwise. 
It is obvious from identities that bidirectional equality can be expressed with \emph{XNOR} (i.e. \emph{NOT XOR}) as given by $(r\leftrightarrow SAT) \equiv [(1-r)\oplus SAT]$.
Therefore, to include an extra bit $r$ for the result of a multi-input \emph{XOR} clause one merely encode the extended
\begin{equation}
(1-r)\oplus \left(\bigoplus\limits_{i=0}^{N-1} x_i\right) \label{eq:r_encoded}
\end{equation}
SAT expression with the strategy described earlier in Sec.~\ref{sec:XORencoding}.
When the multi-input \emph{XOR} defined on the variables $x_i$ ($0\leq i\leq N-1$) evaluates to true (false), (\ref{eq:r_encoded}) can be only satisfied with $r=1$ ($r=0$).

When desiring bidirectional equality for a disjunctive (i.e. \emph{OR}) or conjunctive (i.e. \emph{AND}) formula, a different algorithm will be used.  
Interestingly enough, a conjunctive constraint, due to having only one \emph{min-term}, is directly encode-able in a QUBO form without any substitution variables \cite{10.21203/rs.3.rs-3471221/v1} via 
\begin{equation}
    \left( N - \sum\limits_{i=0}^{N-1} x_i\right)^2\;. \label{eq:simple_sum_QUBO}
\end{equation}
However, this formula contains coefficients of booth small and large magnitude.
This becomes an issue especially in situations when the numerical accuracy becomes limited by the utilized hardware, like D-Waves's annealers.
Here we propose an alternative formula in which the variables appear only on the first power, reducing this way the magnitude of the coefficients.
With even number of variables, the QUBO encoding reads as
\begin{equation}
    \frac{N}{2} - \sum\limits_{i=0}^{N/2-1} x_{2i+1}x_{2i}\;. \label{eq:conjunction_min_coeff}
\end{equation}
For odd $N$ one extra (linear) term is required containing the last variable $x_N-1$ and the constant in the front should be increased by $1$.



We further prove that introducing substitutions for higher order products in \emph{k-DNF} one can simply reuse the existing \emph{k-CNF} method of Ref.\cite{10.1145/3631908.3631929}.  
In this context the terminology of "higher order product" means multi-input conjunctions with at least $3$ inputs.
The \emph{CNF} method in \cite{10.1145/3631908.3631929} is based on the simple transformation similar to the one used in Tseitin encoding to \emph{3-CNF}.  In other words, first we assign a new substitution variable to each of the disjunctive terms in the \emph{CNF} via a strong equivalence.
The introduced substitution variables are then used in a conjunctive relation encode-able via (\ref{eq:conjunction_min_coeff}).
For example, having a \emph{3-CNF} given by
\begin{equation}
    (x\vee y \vee z ) \wedge (x\vee \overline{y} \vee z ) \wedge (\overline{x}\vee y \vee z )
\end{equation}
we assign substitution variables $r_1$, $r_2$ and $r_3$ to the individual disjunctive terms as $r_1\leftrightarrow(x\vee y \vee z )$, $r_2\leftrightarrow(x\vee \overline{y} \vee z )$ and $r_3\leftrightarrow(\overline{x}\vee y \vee z )$. 
Finally, we encode the conjunctive relation $r_1 \wedge r_2 \wedge r_3$ between these substitution variables via Eq.~(\ref{eq:conjunction_min_coeff}).

The strong equivalence of
\begin{equation}
r\leftrightarrow \bigvee\limits_{i=0}^{k-1} x_i \label{eq:strong_equivalence_or}
\end{equation}
can be rewritten by expanding the \emph{XOR} operation as $a \oplus b\equiv (a \vee b)\wedge (\overline{a} \vee \overline{b})$ and making use of DeMorgan's law 
\begin{equation}
    \bigwedge\limits_{i=0}^{k-1} c_i =  \overline{\bigvee\limits_{i=0}^{k-1} \overline{c_i}}\;,
\end{equation}
resulting in 
\begin{equation} 
\left(\overline{r} \vee \bigvee\limits_{i=0}^{k-1} x_i\right) \wedge \left(\bigwedge\limits_{i=0}^{k-1} (r \vee \overline{x_i})\right)\label{eq:r_encoded_or}
\end{equation}
with the assignments $a=\overline{r}$, $b=\bigvee_i x_i$ and $c_i=\overline{x_i}$.
In deriving Eq.~(\ref{eq:r_encoded_or}) we also used the distributive nature of the involved operations, namely that:
\begin{equation}
    r\vee \left(\bigwedge\limits_{i=0}^{k-1}  \overline{x_i}\right) = \bigwedge\limits_{i=0}^{k-1} (r \vee \overline{x_i})\;.
\end{equation}
We interpret Eq.~(\ref{eq:r_encoded_or}) in the following way: on the left side one have a disjunctive clause (or equivalently a multi-input \emph{OR} function) with $k+1$ variables.
On the right side we have a conjunction of $2$-input disjunctive clauses, which is nothing else than a \emph{2-CNF} expression.
Since the clauses in a \emph{2-CNF} formula are directly encode-able via QUBO without introducing further substitution variables, there is only the $(k+1)$-CNF clause to be encoded on the left side of (\ref{eq:r_encoded_or}).
To this end we can use the technique described in Sec.~\ref{sec:ORencoding}.
For encoding the $2$-input disjunctive clauses on the right side $1+rx_i-r-x_i$ can be used.
To encode the conjunction of these terms for all possible $i$ one can sum up these expressions resulting in:
\begin{equation}
    (1-r)\left(k-\sum\limits_{i=0}^{k-1} x_i\right)
\end{equation}
(We notice that the formula (\ref{eq:conjunction_min_coeff}) to encode the conjunctive part of this expression does not work, as it would give cubic terms in the encoding.)
To satisfy the SAT expression (\ref{eq:r_encoded_or}), both sides need to evaluate to true. 
This translates to a unified QUBO encoding as the sum of the QUBO encoding of the left and right sides. 
The QUBO instance takes on its global minimum of $0$, when both components reaches their global minimum at the same time.
In summary, to encode a whole \emph{k-CNF} expression into a QUBO form we make use of the substitution variables introduced by the strong equivalence (\ref{eq:strong_equivalence_or}) for the $k$-term sub-clauses and encode the outer conjunctive relation between all of these variables labeled by $r_p$.
Here, the index $p$ was introduced to label the individual disjunctive terms in the \emph{k-CNF}.
The conjunctive relation between the variables $r_p$ is then encoded via Eq.~(\ref{eq:conjunction_min_coeff}).

The procedure is similar for \emph{k-DNF}, with all the literals in the formula (\ref{eq:r_encoded_or}) negated.
Thus, the strong equivalence  of
\begin{equation}
    r\leftrightarrow \bigwedge\limits_{i=0}^{k-1} x_i \label{eq:strong_equivalence_and}
\end{equation}
introduced for the sub-clauses of the \emph{DNF} form can be written as 
\begin{equation} 
  \left(\bigwedge\limits_{i=0}^{k-1} (\overline{r} \vee x_i)\right) \wedge \left(r \vee \bigvee\limits_{i=0}^{k-1} \overline{x_i}\right) \;. \label{eq:r_encoded_and}
\end{equation}
This equation can be encoded in a QUBO form with the same method as Eq.~(\ref{eq:r_encoded_or}).
Then, the outer multi-input disjunctive relation present in the \emph{DNF} form is encoded with the procedure described in Sec.~\ref{sec:ORencoding} and applied on the substitution variables arising from strong equivalences formulated by Eq.~(\ref{eq:strong_equivalence_and}) for all of the conjunctive sub-terms.

Lastly, we offer a procedure for the \emph{k-ANF} encoding in which the sub-clauses contain $k$ variables.
As the \emph{ANF} form is built up from conjunctive expressions connected with \emph{XOR} operations (like $(x\wedge y)\oplus(\overline{x}\wedge z)$) we might again make use of the substitutions variables introduced for all product terms via Eqs.~(\ref{eq:strong_equivalence_and}) and (\ref{eq:r_encoded_and}), and the outer multi-input \emph{XOR} clause (now implemented on the substitution variables) gets encoded with the parity formulation discussed in Sec.~\ref{sec:XORencoding}.

\section{QUBO Encoding Techniques for Cryptography Functions} \label{sec:crypto}

A simple method of encoding cryptography circuits is to create a well-formed formula (WFF) (i.e. a syntactically correct arrangements of logical symbols and proposition letters in propositional logic) and using Tseitin's encoding to reduce it to \emph{CNF} form.  SAT pre-solvers could potentially further reduce the circuit when in \emph{CNF} form.  This is only as efficient as the expression of the original circuit.  However, there are cases where portions of circuits can be simplified with logic minimizers, which due this problem belonging to NP, will tend to rely on heuristic solutions.  
However, when converting to QUBO form, this is not always convenient.  For example, with any sort of binary addition, it will be a very large chore for a logic minimizer to minimize the circuit when scaling up the problem to large number of terms.  
In turn, a simple sum of variables (like in Eq.~(\ref{eq:simple_sum_QUBO})) which is then squared, is directly encode-able into a QUBO form.  
To avoid such rabbit holes, here we propose a procedure via applying a logic marker (i.e. substitution variables) to some chosen boolean functions $s_{m},\dots,s_1=f(x_{n},\dots,x_1)$ with $n$ inputs and $m$ outputs.  
A logic marker can rely on a custom heuristic.  
Furthermore, we distinguish between logic markers that are going through a logic minimizer and those going directly to QUBO (like the square of a sum mentioned earlier).
During the encoding procedure an issue will be encountered, as Tseitin encoding will not allow the utilization of the developed "multi-input binary gate to QUBO" formulae without logic markers.
To optimize the encoding, multi-input gates must be identified and simplified to fewer variables.  This can be done by intelligent placement or during processing of the WFF while it is being converted to Tseitin (i.e. \emph{CNF}) form.

In principle, there are three generic multi-input gates, which can be put in the focus when searching for optimal QUBO encodings.
These are the \emph{AND}, \emph{OR} and \emph{XOR} gates discussed in previous sections.
The true potential of these gates lies in their commutative and associative properties.  
In a \emph{NOT}-free representation of a WFF rooted at one of these binary logic gate (i.e. the outermost binary operation is one of these gates), the resulting logical expression will made of \emph{IMPLY}, \emph{CIMPLY},  \emph{NIMPLY},  \emph{CNIMPLY} and  \emph{XNOR} operations.  
\begin{table*}
    \centering
    \begin{tabular}{|c|c|c|}
    \hline
    \textbf{Operation} & \textbf{Symbol} & \textbf{Binary equivalent}\\
    \hline
    \emph{NOT} & $\overline{a}$ & $\overline{a}$\\
    \emph{AND} & $a\wedge b$ & $a\wedge b$ \\
    \emph{OR} & $a\vee b$ & $a\vee b$\\
    \emph{NAND} & $a\overline{\wedge} b$ & $\overline{a}\vee\overline{b}$\\
    \emph{NOR} & $a\overline{\vee} b$ & $\overline{a}\wedge\overline{b}$\\
    \emph{XOR} & $a\oplus b$ & $(a\vee b) \wedge (\overline{a}\vee \overline{b})$ \\
    \emph{XNOR} & $a\Leftrightarrow b$ & $(a\vee \overline{b}) \wedge (\overline{a}\vee b)$\\
    \emph{IMPLY} & $a\Rightarrow b$ & $\overline{a}\vee b$\\    
    \emph{NIMPLY} & $a\nRightarrow b$ & $a\wedge\overline{b}$\\    
    \emph{CIMPLY} & $a\Leftarrow b$ & $a\vee \overline{b}$\\    
    \emph{CNIMPLY} & $a\nLeftarrow b$ & $b\wedge\overline{a}$\\    
    \hline
    \end{tabular}
    \caption{Binary equivalents of elementary logical operations. }\label{tbl:elementary}
\end{table*}
In Table.~\ref{tbl:elementary} we summarize the binary equivalent form of the elementary logical operations.
As one can see, all the logical operations can be expressed with the combination of \emph{OR} and \emph{AND} operations.
Therefore, three equivalence classes can be defined depending on which generic operation is used.
The logical operations of the first class encompassing the operations \emph{AND}, \emph{NOR}, \emph{NIMPLY}, \emph{CNIMPLY} can be expressed with conjunction.
The logical operations of the second class encompassing the operations \emph{OR}, \emph{NAND}, \emph{IMPLY}, \emph{CIMPLY} can be expressed with a disjunctive relation.
The third class is made of logical operations \emph{XOR} and \emph{XNOR} having a \emph{CNF} form with two clauses.
Therefore, it is straight forward to express complex logical formulas made up from these operations  via multi-input generic binary operations \emph{OR} and \emph{AND}. 
Notice that these expressions also contain the negated literals of the original formulas. For example, $a\wedge(b\nRightarrow c)$ is expressed via a multi-input conjunction $a \wedge b \wedge \overline{c}$.
When \emph{XOR} is used to express the logical formula in binary format, a constant true value might be needed to involve, ensuring the correctness of the expression, like in the case of $a\Leftrightarrow b$ which translates to $a\oplus b \oplus 1$. 
However, $(a\Leftrightarrow b)\oplus(c\Leftrightarrow d)$ translates to $a\oplus b \oplus c \oplus d$ 
without any constant values, as the two constant true-s coming from \emph{XNOR} operations cancel each other.
A simple depth-first search can thereby simplify any set of equivalence classes and omit constants expressed by false and true, single literals, binary logic gates, or if the reduction is to three or more inputs, a multi-input logic gate substituted by substitution variables. The various cases where literals appear twice or with their negation should be appropriately handled depending on the gate type.
After arriving at a multi-input binary function we can use the developed encoding techniques described in earlier sections. We summarize these encoding formulae in Table \ref{tbl:allform}. 
\begin{table*}
    \centering
    \begin{tabular}{|l|l|}
    \hline
    \textbf{Operation} & \textbf{QUBO encoding formula}\\
    \hline
    Multi-input XOR: $\bigoplus_i x_i$  & Eq.~(\ref{eq:parity_QUBO})\\
    Popcount range: $L\le \sum_i x_i \le H$  & 
    Eqs.~(\ref{eq:general_multi_input_QUBO}), (\ref{eq:f1_general}) and (\ref{eq:f2_general})\\
    Multi-input OR: $\bigvee_i x_i$  & 
    Eqs.~(\ref{eq:general_multi_input_QUBO}), (\ref{eq:f1_general}) and (\ref{eq:f2_general}) with $L=0$ and $H=N$\\  
    Multi-input NAND: $\overline{\bigwedge} X$  & Eqs.~(\ref{eq:general_multi_input_QUBO}), (\ref{eq:f1_general}) and (\ref{eq:f2_general}) with $L=0$ and $H=N-1$ \\
    Multi-input AND: $\bigwedge_i x_i$  & Eq.~(\ref{eq:simple_sum_QUBO}) or (\ref{eq:conjunction_min_coeff})\\ 
    \hline
    \end{tabular}
    \caption{Summary of encoding strategies for multi-input binary functions.}\label{tbl:allform}
\end{table*}


\subsection{Advanced Encryption Standard (AES)}

A natural application that arises containing an abundance of \emph{XOR} constraints is the Advanced Encryption Standard (AES) \cite{10.1007/10721064_26,daemen02} which beyond the non-linear affine transformations inside of the substitution boxes (S-boxes), consists entirely of \emph{XOR} operations, or with other words, additions in Galois field with two elements (GF($2$)).  This is thematic throughout many symmetric cryptography schemes, in fact.

AES is a cryptography scheme using a symmetric key, that is the same key is used to both decoding and encoding.
The key size used for an AES cipher specifies the number of transformation rounds that convert the input into the final encoded output. The number of rounds are 
$10$ rounds for 128-bit keys, $12$ rounds for 192-bit keys and $14$ rounds for 256-bit keys.
The individual rounds break down to $4$ steps performed on an input of $4\;\textrm{byte}\times4\;\textrm{byte}$ data blocks of the input: (i) substituting each of the bytes via a substitution box (S-box), (ii) shifting rows in the $4\;\textrm{byte}\times4\;\textrm{byte}$ matrix, (iii) mixing columns in this matrix and finally, (iv) add a cipher key to the resulting elements of the matrix with bit-wise \emph{XOR} operation. 
In each round the output of the previous round is processed with a newly generated $16$ byte key ((iv)-th step) obtained from the initial AES key (which is either $128$, $192$ or $256$ bit long) via a key schedule technique.
At the beginning, before performing any rounds, an instance of step (iv) is applied on the input.
The last round omits step (iii).
In order to recover the initial key and decode the message, one might follow the route of translating the AES transformation rounds into a boolean circuit (i.e. a circuit composed from boolean functions like the \emph{AND}, \emph{OR} or \emph{XOR} gates) which can be further processed into a QUBO form using the developed techniques in this work.
During this procedure the bits of the key are considered to be unknown, while the bits of the input text and the encrypted data are known constants.
The global minimum of the QUBO will satisfy the boolean circuit and the encryption key can be recovered.
For AES, the S-box represents the additional challenge for optimal QUBO encoding.
This could be achieved by designing an efficient boolean circuit of the multiplication inverse function over GF($2^8$) and than encode it into a QUBO form. 
However, searching for an optimal boolean circuit representation is an NP hard problem in itself \cite{cryptoeprint:2020/530}.
An efficient decomposition of the multiplication inverse over GF($2^8$) was given by Canright in \cite{10.1007/11545262_32}.  
According to these results, the multiplication inverse in GF($2^8$) can be expressed via $3$ multiplications in GF($2^4$) and a single multiplication inverse in GF($2^4$).
Additionally, a GF($2^4$) multiplication is decomposed in terms of $3$ GF($2^2$) multiplications.
Finally, the multiplication in GF($2^2$) is reduced enough in complexity to be expressed via optimal boolean circuit.
The \emph{AND} gates in the circuit of the multiplication appear minimally three times, defining the multiplicative complexity of this operation to $3$.
Three GF($2^2$) multiplications also occur in one use of the inverse modulo in GF($2^4$) as well.
In summary, the boolean circuit of a GF($2^8$) has a complexity of $3\times3\times3+3\times3=36$ \emph{AND} gates.
All of the inputs to these gates require \emph{XOR} substitutions (described in Sec.~\ref{sec:subst_encoding}), though there are several which are re-usable. However, we note an observation similar to that of Boyar and Peralta \cite{10.1007/978-3-642-30436-1_24} that the multiplication inverse of GF($2^4$) has an exhaustively provable multiplicative complexity of $5$ \emph{AND} gates which is a decent savings from the original $3\times 3=9$ \emph{AND} gates.
(The exhaustive search is carried out with the procedure described in Sec.~\ref{sec:exhaustive_search} via the polynomial $g$, the \emph{AND} gates in $g$ are equivalent to quadratic multiplications. We discuss the details of this procedure later in this section.) 
In fact, to date $3\times3\times3+5=32$ is the best known result for the number of non-\emph{XOR} gates required in the circuit of the GF($2^8$) multiplication inverse circuit as reported more recently in \cite{Reyhani-Masoleh_Taha_Ashmawy_2018}.  
However, it remains unclear if multiplication in GF($2^4$) and multiplication inverse in GF($2^8$) have optimal multiplicative complexities of $9$ and $32$ if expressed via boolean circuits, or less.
This is because determining the multiplicative complexity of boolean functions is considered intractable beyond $6$ variables \cite{cryptoeprint:2020/530}.
In line with this finding, we experienced these problems to be extremely challenging in the context of our exhaustive search technique presented in  Sec.~\ref{sec:exhaustive_search} and failed to tackle either of them.
Any remaining details in QUBO encoding of AES (steps (ii)-(iv) in the individual rounds) involve a reuse and simplification of the \emph{XOR} formulae developed in Sec.~\ref{sec:XORencoding}. 
In order to showcase the potential in optimizing a set of multi-input \emph{XOR} formulae we outline the following example.
Having two \emph{XOR} expressions $z_0\leftrightarrow (a\oplus b\oplus c\oplus d$ and $z_1\leftrightarrow (a\oplus b\oplus c\oplus e)$ we can reuse $z_0$ in the second expression as $z_1\leftrightarrow (z_0\oplus d\oplus e)$ and reduce the expression in input variable count. 
To encode a shorter multi-input $\emph{XOR}$ expression we need to cover a smaller range with the substitution variables, hence saving variables in the encoding.
As the optimal reduction of these \emph{XOR} formulae is a parity-style variant of the NP-hard minimum set covering problem, we opted a heuristic using a greedy algorithm to first find the largest reusable $\emph{XOR}$ substitution, then the next largest one, etc.
We also placed substitution variables on all variables at the input to the multiplication of GF($2^8$)
beyond the substitutions for the involved functions of multiplication and associated multiplication inverse in GF($2^4$). 
It should be noted that on order of hundreds to thousands of \emph{XOR} counting set substitution variables can be saved by careful heuristics.

Now we discuss in more details the procedure to solve the QUBO encoding problem of the multiplication function in GF($2^2$) and multiplication inverse in GF($2^4$) needed to encode the S-box.
The multiplication function in GF($2^2$) calculates the product of two elements in GF($2^2$), both described via two bits. 
Therefore the multiplication maps from two bit-pairs onto a single bit-pair: $(x_1,x_0, y_1,y_0)\rightarrow(z_1,z_0)$.
The function is fully represented via the look-up-table between the input-output bits.
The $\mathbf{x}$ bit-string (first input in the polynomial $g$ given by Eq.~(\ref{eq:general_g})) has $6$ elements. 
The bit combinations in the look-up-table determine the \emph{min-terms} for which the QUBO encoding should give the global minimum $0$.
For any other bit-string combination the polynomial $g$ needs to be greater than $0$.
To find the suitable coefficients we solve the ILP problem given by Eq.~(\ref{eq:ILP-full}).
The ILP could be solved with two substitution variables $s_1$ and $s_2$, giving the following QUBO encoding: 
\begin{equation}  
\begin{split}
4x_0+3z_0+17z_1-4x_0x_1-3x_0y_1-4x_0z_0+6x_0z_1\\
+1x_1y_0+4x_1y_1+2x_1z_0-7x_1z_1-2y_0z_0-10y_0z_1\\
+2y_1z_0-5y_1z_1-4z_0z_1+8s_0x_0-7s_0x_1+11s_0y_0\\
-7s_0y_1-8s_0z_0-15s_1x_0+13s_1x_1+5s_1y_0+9s_1y_1 \\
+10s_1z_0-18s_1z_1+11s_0+11s_1-17s_0s_1\;.
\end{split}
\end{equation}
The substitution variables do not distinguish uniquely the combinations of $\mathbf{x}$, thus we are speaking about weak substitution. 
According to which of the $\tilde{g}(\mathbf{x}_{\textrm{min}}, \mathbf{s}_i)$ ($i\in\{0,1,\dots,2^{N_s}-1\}$) terms in Eq.~(\ref{eq:ILP2}) satisfy the ILP, we deduced the substitution variables to be expressed by:
\begin{equation}   
\begin{split}
s_0\;\leftrightarrow\; & y_0\wedge \overline{y_1}\wedge (x_0\vee x_1)\wedge (\overline{x_1}\vee \overline{ z_0}\vee \overline{z_1}) \\
&\wedge (x_0\vee z_0)\wedge (\overline{x_0}\vee z_1) \wedge (x_1\vee z_0)
\end{split}
\end{equation}
and
\begin{equation}  
\begin{split}
s_1\;\leftrightarrow\; & z_1\wedge (\overline{x_1}\vee \overline{z_0})\wedge (\overline{y_1}\vee \overline{ z_0})\wedge (\overline{x_0}\vee y_0\vee y_1)\\
&\wedge (x_1\vee \overline{y_0}\vee z_0)\wedge (x_0\vee \overline{z_0})\wedge (x_0\vee \overline{x_1}) \\
&\wedge (\overline{x_1}\vee \overline{y_1})\wedge (x_0\vee \overline{y_1})\;.
\end{split}
\end{equation}
To encode the multiplication inverse of GF($2^4$),
we follow a similar procedure. 
In this case we are dealing with a map with a single four-bit input and a four-bit output: $(x_1,x_0,x_2,x_3)\rightarrow(z_3,z_2,z_1,z_0)$.
Therefore, the bit-string $\mathbf{x}$ (again, $\mathbf{x}$ labels the input of the polynomial $g$) has $8$ elements, for which we could solve the associated ILP problem with a single substitution variable $s$, giving a QUBO form: 
\begin{equation}    
\begin{split}
    & 32x_0-13x_1+31x_2+73x_3-3z_1-17z_3+21x_0x_1  \\ 
    & -49x_0x_2-32x_0x_3+65x_0z_0 +10x_0z_1-59x_0z_2 \\
    & -38x_0z_3-13x_1x_2 -19x_1x_3+26x_1z_0+4x_1z_1 \\ 
    & -9x_1z_2 +2x_1z_3+26x_2x_3-58x_2z_0-12x_2z_1 \\ 
    & +38x_2z_2+18x_2z_3-59x_3z_0-8x_3z_1-20x_3z_3 \\ 
    & +14z_0z_1-27z_0z_2-6z_1z_2+54z_2z_3+32sx_1 \\ 
    & -72sx_3+59sz_0+10sz_1 +59sz_2+75sz_3-28s+28\;.
\end{split}
\end{equation}
The substitution variable fulfills the relation:
\begin{equation}   
\begin{split}
s\;\leftrightarrow \; &\overline{x_0}\wedge \overline{x_1}\wedge \overline{z_2}\wedge \overline{z_3}\wedge (\overline{x_2}\vee \overline{z_1})\wedge (\overline{x_2}\vee z_0) \\
&\wedge (z_0\vee \overline{z_1})
\wedge (x_2\vee x_3\vee \overline{z_0})\wedge (\overline{x_3}\vee z_1)\;.
\end{split}
\end{equation}
We notice, that our results, due to the exhaustive search, are optimal in the number of substitution variables.
\begin{table*}
\centering
\begin{tabular}{|l|c|c|c|}
\hline
Function & QUBO Size & Density & Max Abs. Coefficient\\
\hline
AES-128 Encryption & 21294 & 0.22\% & 2485  \\
AES-128 Decryption & 22802 & 0.52\% & 15410  \\
AES-192 Encryption & 24264 & 0.21\% & 3446 \\
AES-192 Decryption & 25959 & 0.48\% & 19857 \\
AES-256 Encryption & 29330 & 0.15\% & 2245 \\
AES-256 Decryption & 31242 & 0.34\% & 13935 \\
MD5$^a$ & 10272 & 0.37\% & 18752 \\
MD5$^b$ & 15320 & 0.09\% & 55 \\
SHA1$^a$ & 15587 & 0.27\% & 62600 \\
SHA1$^b$ & 19426 & 0.09\% & 55 \\
SHA-256$^a$ & 30255 & 0.17\% & 66000 \\
SHA-256$^b$ & 42899 & 0.04\% & 55 \\
\hline
\end{tabular}
\caption{The number of variables in the QUBO encoding of several Cryptography Functions. The results also indicate the sparseness of the resulting QUBO matrix (the density is calculated as a ratio of the nonzero elements in the QUBO matrix and $N^2$, with $N$ being the number of variables), and the magnitude of the largest element in the matrix.
The encoding results for MD5, SHA1 and SHA-256 are presented for $2$ cases: (a) using blocks of size $B=6$ for modular addition and (b) $B=1$ with \emph{k-SAT}/\emph{ANF} forms partitioned into at most $13$ clauses (to keep the coefficients low), see the main text for explanation. For encoding with $B=6$ and for \emph{AES} there was no limitation on the length of the SAT forms.} \label{tbl:encres}
\end{table*}
To design the final QUBO representation of the AES encoding, we sum up the QUBOs associated with the individual multiplications and inverse multiplications in GF($2^2$) and GF($2^4$) associated with step (i) of the AES round and also QUBO-encode all the remaining steps of (ii), (iii) and (iv) (composed from \emph{XOR} operations) and add them to the QUBO.
The individual QUBOs are linked via their input variables: if a given computational step depends on the result of a preceding step, then the QUBO variables encoding the inputs of this step overlap with the QUBO variables encoding the output variables of the preceding step.
This provides a straightforward procedure to construct the QUBO representation of the AES encoding.

In Table.~\ref{tbl:encres} we summarize our results on QUBO encoding of AES-128, AES-192 and AES-256,  for both encryption and decryption. 
When comparing our results to prior QUBO encodings of AES reported in \cite{9690063,Burek_Mank_Wronski_2023,a15020033}, we find our results significantly outperforming them.
Firstly, following the work of Ref.~\cite{9690063} one ends up with $30026$ QUBO variables when encoding a single S-box for AES-128, while the whole AES-128 is composed from $10$ rounds, each involving an S-box transformation.
For AES-192 and AES-256, a single S-box requires $58920$ and $70059$ QUBO variables. 
In both cases these numbers turned to be larger than the size of the overall QUBO matrix encompassing all of the $12$ and $14$ rounds of S-box iterations in our approach, respectively.
Even if the S-box encoding was improved by \cite{Burek_Mank_Wronski_2023} to $29528$ (AES-128),  $57384$ (AES-192) and $68187$ (AES-256), the resulting QUBOs would still highly overestimate the problem size required to represent these problems with a QUBO formalism.
Finally, let us mention the encoding strategy reported in \cite{a15020033}, giving $\sim90000$ QUBO variables for AES-128, $\sim200000$ for AES-192 and $\sim250000$ variables for AES-256.
While these results are the most efficient QUBO encoding of AES reported so far, our QUBO encoding requires only the $24\%$, $12\%$ and $12\%$ of the QUBO variable count to encode the whole AES-128, AES-192 and AES-256 problem, respectively.
We provide examples on the QUBO-encoded AES problems via a GitHub repository accessible at \cite{QUBO_github}.

We believe that further improvement can be made upon the AES encoding if one could solve either the optimal boolean circuit for the multiplication in GF($2^4$) or GF($2^8$) or the multiplication inverse in GF($2^8$).
We leave this possibility for future studies, as our exhaustive search did not solve these problems.

\subsection{Encoding constructions of MD5, SHA-1 and SHA-256 hash functions} \label{sec:crypto_examples}

A hash function is a mathematical algorithm that takes a variable-length input, such as a string or a binary data, and generates a fixed-length output, known as a hash value.
The key properties of a good hash function are that it is deterministic, meaning that the same input always produces the same output, and that it is non-invertible, meaning that it is computationally infeasible to recover the original input from the output. Hash functions are widely used in various applications, including data integrity, data compression, and cryptography, to ensure the authenticity and integrity of data.

Among others, the MD5 message-digest algorithm is a widely used hash function producing a $128$-bit hash value. 
MD5 was designed by Ronald Rivest in 1991 to replace an earlier hash function MD4 \cite{rfc1321}.
Initially it was widely used as a cryptography hash function, until it has been found to suffer from extensive vulnerabilities \cite{cryptoeprint:2006/104,md5attack}.
Though, it remains suitable for other non-cryptography purposes, and may be preferred due to lower computational requirements than the more recent Secure Hash Algorithms \cite{Handschuh2005}.
(For example, MD5 can be still used as a checksum to verify data integrity against unintentional corruption.)
The Secure Hash Algorithms are a family of cryptography hash functions including: 
\begin{itemize}
    \item \emph{SHA-1} is $160$-bit hash function resembling the earlier MD5 algorithm. However, cryptography weaknesses were discovered in SHA-1 as well, and the standard was no longer approved for most cryptography uses after 2010.
    \item \emph{SHA-2} is a family of two similar hash functions with different block sizes, known as SHA-256 and SHA-512. While we use SHA-256 to generate $32$-bit words, SHA-512 is outputting $64$-bit words. 
\end{itemize}
As the main result of our work is the QUBO encoding of the AES we omit the detailed mathematical description of these hash algorithms. 
We just mention that they rely on a modular addition over GF($2^{32}$) as well as $3$ and $4$-input \emph{XOR} operations, ternary conditional functions (ternary conditional function operates on $3$ inputs: one binary input is used to choose between the two remaining inputs.) and majority conditional functions (MCF), like the one defined on $3$ input variables: $MCF(x, y, z) = (x \wedge y) \vee (y \wedge z) \vee (z \wedge x)$.
In other words, the MCF returns $1$ if at least two of the input variables are $1$, and $0$ otherwise.
The key point in encoding the mentioned hash functions is the efficient encoding of these components.
The $4$-input \emph{XOR} (used in SHA-256) with substitution variable encoding its result (see Eq.~(\ref{eq:XOR_strong_equivalence})) can be encoded with the procedure described in Sec.~\ref{sec:subst_encoding}.
For the $3$-input \emph{XOR} formula $z\leftrightarrow x_0\oplus x_1\oplus x_2$ we provide an alternative QUBO encoding with a single substitution variable $(x_0+x_1+x_2-z+2s-2)^2$ obtained by ILP.
The weak substitution variable takes on the value $1$ and minimizes the formula when $x_0+x_1+x_2+1-z=1$.
The ternary selection $z\leftrightarrow(x_0\wedge x_1)\vee(\overline{x_0}\wedge x_2)$ is encoded via a QUBO form:
\begin{equation}
    \begin{split}
        &-2x_0-2x_1-1x_2+5z+3x_0x_1+1x_0x_2-2x_0z+2x_1x_2 \\
        &-4x_1z-2x_2z+4sx_0+6sx_1+2sx_2-4sz-2s+2\;.
    \end{split}
\end{equation}
The encoding obtained by solving the associated ILP resulted in a single substitution variable $s$.
It takes on the value $1$ when $x_0+x_1+x_2+1-z=1$.
Additionally, the $3$-input majority function $z\leftrightarrow x_0 \wedge x_1 \vee x_0 \wedge x_2 \vee x_1 \wedge x_2$ can be encoded without any substitution variable via the QUBO form 
\begin{equation}
    3z+1x_0x_1+1x_0x_2-2x_0z+1x_1x_2-2x_1z-2x_2z\;.
\end{equation}
We also need to mention one last function when QUBO-encoding the MD5 hash function.
This is the auxiliary formula $z\leftrightarrow x_0 \oplus (x_1 \vee \overline{x_2})$ included in the definition of MD5 and encoded via:
\begin{equation}
    \begin{split}
        &-7x_0-8x_1-5x_2-7z+4x_0x_1+2x_0x_2+4x_0z+3x_1x_2 \\
        &+4x_1z+2x_2z+6sx_0+8sx_1+4sx_2+6sz-10s+11\;.
    \end{split}
\end{equation}
The substitution variable $s$ takes on the value $1$ and minimizes the formula when $\overline{x_1} \wedge (x_0+x_2+x_3 \overset{?}{=} 1)$ is True.
Finally, only the discussion to encode the modular addition remains.
Generally speaking, the most efficient way to encode addition in a QUBO is to write the addition as an equality $a+b=c$, then squaring the difference of the left and right sides and take the minimum of the quadratic function: $(a+b-c)^2\rightarrow0$.
When expanding $a$, $b$ and $c$ into binary formats, we obtain the QUBO representation of the addition. 
The drawback of this simple approach are the large coefficients in front of the most significant bit-variables.
When dealing with $32$-bit addition relevant in the case of the studied hash functions, these coefficients become to large to be applicable on QUBO solving hardware.
To address this issue, Ref.~\cite{jiang2018quantum} developed a block-wise strategy to break the QUBO encoding of additions into smaller blocks.
The QUBO formulae of the blocks are inter-connected via carry bits, as the sum of the terms might overflow the range covered by the bits of the block.
Following this concept, we split the inputs into blocks by
\begin{equation}
    a = \sum\limits_{i=0}^{i_{max}-1} a_{i} 2^{(iB)}, 
\end{equation}
where $B\geq1$ ($B\in\mathbb{N}$) is the block-size. 
With other words, the value of $B$ tells us the number of "ordinary" bits encompassed via a single coefficient $a_i$ ($0\leq i \leq i_{max}-1$ with $i_{max}=\lceil\log_2(a)\rceil/B$, given that $B$ divides $\lceil\log_2(a)\rceil$ with no remainder) in a range $0\leq a_i\leq 2^B-1$.
By expanding $b$ and $c$ as well, the addition reads as
\begin{equation}
    0 = \sum\limits_{i=0}^{i_{max}-1} \left(a_{i} + b_i - c_i\right) 2^{(iB)}. \label{eq:partitioned_add}
\end{equation}
At this point, however, we need to account for potential overflows in the additions $a_{i} + b_i - c_i$ as none of the coefficients is allowed to get larger than $2^B-1$.  
This can be accomplished by introducing carry (or substitution) variables labeled by $s_i$, encoding the potential overflow of the sub-addition $a_{i} + b_i$.
With this logic Eq.~(\ref{eq:partitioned_add}) is equivalent to the formulation
\begin{equation}
    \begin{array}{ll}
        0 = a_{0} + b_0 - c_0 - 2^Bs_1 & \textrm{if: } i=0\\
        0 = a_{i} + b_i - c_i + s_{i} - 2^Bs_{i+1} & \textrm{if: } 0< i\leq i_{max}-1\;.
    \end{array} \label{eq:partitioned_modular_add}
\end{equation}
The last substitution variable $s_{i_{max}}$ ensures that the addition remains always valid, as $c_{i_{max}-1}$ can not exceed $2^B$.
This also ensures the modular nature of the addition with the modulus $2^{i_{max}B}$.
When encoding the addition in QUBO form, one need to replace $a_i$, $b_i$ and $c_i$ with their binary expansions ($s_i$-s are already binary), square the individual rows and sum them up. 
This way we can encode the modular addition in GF($i_{max}B$), with maximal coefficients of $2^{2B}$ due to the squaring of the binary expressions.

In the MD5 algorithm the modular additions can be grouped into a single four-input and one two-input addition over 32-bit numbers in each round.
The addition with $4$ inputs implies the modification of (\ref{eq:partitioned_modular_add}) to
\begin{equation}
    \begin{array}{ll}
        0 = a_{0} + b_0 + c_0 + d_0 - e_0 - & \textrm{if: } i=0 \\
        - 2^Bs_0 - 2^{B+1}s_1\ & \\
        0 = a_{i} + b_i + c_i + d_i - e_i + s_{2i-2} + &  \textrm{if: } 0< i\leq i_{max}-1 \\
         + 2s_{2i-1} -  2^Bs_{2i} - 2^{B+1}s_{2i+1}  &  
    \end{array} \label{eq:partitioned_modular_add4}
\end{equation}
describing the four-term addition $a+b+c+d = e$. 
The $4$ terms in the addition imply $2$ extra bits in each block to cover the potential overflow in the addition. 
The lower bound of maximal coefficient in the QUBO encoding due to expanding the squares lead to $2^{2B+2}$.
We have also tried to further simplify the encoding logic with the ILP strategy developed in Sec.~\ref{sec:exhaustive_search}, though did not find any better encoding with less variables.
In the actual MD5 implementation some pre-computed constants are also used in the procedure opening the door for possible reductions in the variable count by substitution constant bits and merging QUBO terms. 
Because of this, the coefficient in front of the merged QUBO terms are expected to further increase by some extent.
We summarize our results on MD5 QUBO encoding in Table.~\ref{tbl:encres}. 
The encoding with $B=6$ (MD5$^a$) seems to be more efficient than with $B=1$ (MD5$^b$). 
This is because the $B=1$ case involves roughly $6$-times more carry bits than the encoding with $B=6$ sized blocks.
Also, a limit on the maximal length of \emph{ANF} forms (i.e. the maximal number of inputs in \emph{XOR} operations) has been set to $k=26$ for the case of $B=1$. (For the $B=6$ case and for the \emph{AES} QUBO encodings we left the length of the SAT forms unlimited.)
This little adjustment can be also used to decrease the coefficients in the QUBO instance, as the substitution variables would need to cover smaller ranges. 
This good practice becomes especially valuable when needed to port the QUBO instance onto a machine with limited numerical precision.
Longer \emph{ANF} forms are partitioned into smaller segments via substitution variables according to the procedure in Sec.~\ref{sec:subst_encoding}.
Finally, we need to mention that savings in the variable count usually comes with the price of larger coefficients, as reported in Table.~\ref{tbl:encres}.
To run calculations on an actual hardware a reasonable trade-off needs to be found between QUBO variable count and the magnitude of the coefficients.
Comparing to the QUBO encoding of Ref.~\cite{a15020033} having $17280$ QUBO variables, our encoding strategy brings in a significant reduction to $10272$ QUBO variables obtained with $B=6$ sized blocks for addition.

We also provide QUBO instances on larger hash function like the SHA1 and SHA-256.
The main difference compared to MD5 in relation to QUBO encoding are the implemented modular additions.
While MD5 incorporates one addition over $4$ inputs and one addition on $2$ inputs, SHA involves additions that can be joined into a single addition with $5$ inputs, while the additions in SHA-256 can be grouped into additions with $6$ and $7$ inputs.
To encode these addition into a QUBO form, we further modify Eq.~(\ref{eq:partitioned_modular_add4}). 
Aside having more inputs, we also need to introduce one more carry bit in each block accounting for possible overflows between the blocks. 
Because of the extra carry bit, the magnitude of the maximal elements in the QUBO matrix gets increased to $\sim 2^{2B+4}$ as reported in Table.~\ref{tbl:encres} for the encoding of SHA1$^a$ and SHA-256$^a$ for $B=6$.
Comparing to the QUBO encoding reported in Ref.~\cite{a15020033} on SHA-256, we provide a QUBO encoding with $30255$ variables compared to $47808$ variables of Ref.~\cite{a15020033}.
This was mainly achieved by partitioning the modular additions into blocks of size $B=6$.
When targeting QUBO instances with smaller elements, we can use blocks with $B=1$, still improving on the result of Ref.~\cite{a15020033} with $42899$ QUBO variables and maximal elements of $55$.
Our results can be accesses via GitHub repository at Ref.~\cite{QUBO_github}.

\section{Conclusions} \label{sec:conclusions}

In summary, in this work we have reported on a generic method to construct compact QUBO encodings of computational problems with reduced number of binary variables.
We have developed variable-efficient encoding for several generic patterns, such as \emph{CNF}, \emph{DNF},\emph{ANF} and \emph{PNF} SAT forms.
While substitution variables can be introduced for the inner operators following the idea of Sec.~\ref{sec:subst_encoding}, the outer multi-input operation is encoded with one of the strategies reported in Sections.~\ref{sec:XORencoding} for \emph{XOR}, \ref{sec:ORencoding} for \emph{OR} and  \ref{sec:subst_encoding} for \emph{AND} function.  
The key moment in these strategies is the Root Squeezing Theorem enabling us to reduce the number of the substitution variables to the minimum. The optimality of these encodings was proven via exhaustive ILP search on smaller instances.

As generic building blocks, these patterns can be used as building blocks to construct the QUBO encoding of computational problems.
Specific functions in these constructions, like the modular multiplication and modular addition in GF($2^n$), can always be encoded by the ILP strategy of Sec.~\ref{sec:exhaustive_search} to find the optimal QUBO form. 
Though, solving the associated ILP problem is quite challenging, thus breaking these special function into smaller components is inevitable. 
This is what happened in the encoding of the of the multiplicative inverse in GF($2^8$) used to generate the QUBO form of the AES, by breaking the problem into smaller arithmetic operations.
The QUBO instances generated this way are not optimal in the variable count, as additional substitution variables needs to be introduced to convey intermediate results.
Though, as we have seen in relation with our results in Table.~\ref{tbl:encres} obtained for several cryptography constructions, the outlined encoding procedure turned to be extremely powerful.
Our encoding strategy significantly improved on prior results in all of the addressed use cases of AES-128/192/256 encryption schemes and MD5, SHA1 and SHA-256 hash functions. 
(Even if MD5 and SHA1 turned to be vulnerable against cyber attacks \cite{cryptoeprint:2006/104,md5attack,10.1007/978-3-319-63688-7_19,cryptoeprint:2014/239}, they are more than good candidates to examine the developed encoding strategy due to their complexity.)
In particular, for AES-256 our QUBO encoding took only $12\%$ of the size of previously published smallest encoding of Ref.~\cite{a15020033}.
Our results covey the QUBO encoding of AES-256 with $\sim30000$ QUBO variables, which is still beyond the scope of current capabilities of QUBO solving machines (in cryptography applications only the  exact global minimum can be accepted as a solution).
However, the possibility to construct compact QUBO instances for generic computational problems like the ones studied in this work, clearly highlights the potential in systems designed to solve QUBO problems.
The degree of the sparsity ($0.04\%-0.52\%$) of the generated QUBO instances is also notable, as it favours for the embedding onto limited connectivity hardware, like quantum annealers.

However, there remains work to be done in finding a more efficient ILP method to find the QUBO encoding of specific functions expressed with general boolean formulas.
Whether this is by a construction with increased convexity, or via an objective function which guides towards feasibility and helps identify cutting planes.
This way the encoding results are expected to further reduce in size, which is a critical aspect in porting optimization problems onto QUBO solving hardware, like quantum annealers.
In relation with these machines, we need to mention their limitations in numerical precision, such that the elements in the QUBO matrices need to be fit into a given bit-width representation.
To reduce the magnitude of the coefficients of the QUBO terms we introduce a good practice of limiting the maximal size of QUBO encoding patterns.
While the coefficients gets smaller due to the reduced range covered by substitution variables, the price for achieving this becomes the increase in the variable count, as additional substitution variables are needed to encode the intermediate results.

The most important limitations or unknown aspects of QUBO encoding is when dealing with general binary functions constructed for special functions. In circuit minimization and optimization, though finding satisfiable possibilities scales even higher, unsatisfiability typically becomes intractable at $6$-variable functions, due to the practical explosion in the search space \cite{cryptoeprint:2020/530}.  While techniques like symmetry breaking, slack variables, guided counter-example optimization, and other common reductions help, these improvements generally have only served to make e.g. $6$-variable functions tractable for many contexts.  With ILPs, the situation is very similar as expected.
As finding the optimal encoding for larger boolean functions (like the sum-of-products formulae and integer multiplication) would significantly improve on the variable count in QUBO encodings, the study on further improvement of the general ILP formulation of binary functions and solving these problems deserves  significant attention.
We leave these studies for a future work. Future patent applications will be filed to further develop and expand the encoding techniques described in this paper.

\section*{Acknowledgments}

In this manuscript, we would like to acknowledge the valuable contributions of Daniela Theis and Ernesto Bonomi, who reviewed our manuscript and provided insightful comments and suggestions. Their feedback has significantly improved the quality and clarity of our work, and we are grateful for their time and expertise.

\bibliographystyle{elsarticle-num-names} 
\bibliography{biblio}
\end{document}